\documentclass[a4paper,showpacs,preprintnumbers,amsmath,amssymb,pra,superscript]{revtex4}
\usepackage{amsopn}
\usepackage{amsmath}
\usepackage{amssymb}
\usepackage{hyperref}
\usepackage{graphicx}
\usepackage{color}
\usepackage{listings}
\usepackage{todonotes}
\usepackage{tikz}
\usepackage{subfig}
\usetikzlibrary{arrows,decorations.pathreplacing}

\newcommand{\ket}[1]{\ensuremath{|#1\rangle}}
\newcommand{\bra}[1]{\ensuremath{\langle#1|}}
\newcommand{\ketbra}[2]{\ensuremath{\ket{#1}\bra{#2}}}

\newcommand{\HH}{\mathcal{H}}
\newcommand{\Tr}{\mathrm{Tr}}
\newcommand{\tr}{\Tr}

\newcommand{\1}{{\rm 1\hspace{-0.9mm}l}}

\newcommand{\ii}{\mathrm{i}}
\newcommand{\dd}{\mathrm{d}}

\newcommand{\LL}{\mathcal{L}}
\newcommand{\GG}{\mathcal{G}}
\newcommand{\FF}{\mathcal{F}}
\newcommand{\BB}{\mathcal{B}}
\newcommand{\CC}{\mathcal{C}}
\newcommand{\A}{\mathcal{A}}

\begin{document}

\title{Various methods of optimizing control pulses for quantum systems with 
decoherence}

\author{{\L}ukasz Pawela}
\email{lpawela@iitis.pl}
\affiliation{Institute of Theoretical and Applied Informatics, Polish Academy
of Sciences, Ba{\l}tycka 5, 44-100 Gliwice, Poland}

\author{Przemys{\l}aw Sadowski}
\email{psadowski@iitis.pl}
\affiliation{Institute of Theoretical and Applied Informatics, Polish Academy
of Sciences, Ba{\l}tycka 5, 44-100 Gliwice, Poland}

\begin{abstract}
We study three methods of obtaining an approximation of unitary evolution of a 
quantum system under decoherence. We use three methods of optimizing the 
control pulses: genetic optimization, approximate evolution method and 
approximate gradient method. To model the noise in the system we use the 
Lindblad equation. We obtain results showing that genetic optimization may 
give a better approximation of a unitary evolution in the case of high noise.
\end{abstract}

\date{21/VII/2014}



\maketitle

\section{Introduction}\label{sec:introduction}
One of the fundamental issues of quantum information science is the ability
to manipulate the dynamics of a given complex quantum system. Since the
beginning of quantum mechanics, controlling a quantum system has been an
implicit goal of quantum physics, chemistry and implementations of quantum
information 
processing~\cite{cheng2005optimal}.

If a given quantum system is controllable, i.e. it is possible to drive it into
a previously fixed state, it is desirable to develop a control strategy to
accomplish the required control task.  In the case of finite dimensional 
quantum
systems the criteria for  controllability can be expressed in terms of
Lie-algebraic concepts
\cite{albertini2002lie,elliott2009bilinear,d2008introduction}. These concepts
provide a mathematical tool, in the case of closed quantum systems, i.e. 
systems
without external influences. 

It is an important question whether the system is controllable when the
control  is performed only on a subsystem. This kind of approach is called a
\emph{local-controllability} and can be considered only in the case when the
subsystems of  a given system interact. As examples may serve coupled spin
chains or spin networks 
\cite{d2008introduction,burgarth2007full,burgarth2009local,puchala2012local}.
Local-control has a practical importance in proposed quantum computer
architectures, as its implementation is simpler and the effect of decoherence
is reduced by decreased  number of control
actuators~\cite{montangero2007robust,fisher2010optimal}.

A widely used method for manipulating a quantum system is a coherent control
strategy, where the manipulation of the quantum states is achieved by applying
semi-classical potentials in a fashion that preserves quantum coherence. In the
case when a system is controllable it is a point of interest what actions must
be performed to control a system most efficiently, bearing in mind limitations
imposed by practical 
restrictions~\cite{morzhin2012nonlocal,dominy2013analysis,qi2013two,pawela2012quantum,pawela2013quantum}.
The always present noise in the quantum system may be considered as a such
constrain~\cite{li2009pseudospectral,khodjasteh2010arbitrarily,schmidt2011optimal,gawron2013decoherence,schulte2011optimal,hwang2012optimal,clausen2012task,floether2012robust}.
Therefore it is necessary to study
methods of obtaining piecewise constant control pulses which implement the
desired quantum operation on a noisy system.

In this paper we present a general method of obtaining a piecewise-constant
controls, which is robust with respect to noise in the quantum system. 
This means, we
wish to perform a unitary evolution on a greater system than the target one, 
and discard the ancilla.

This paper is organized as follows. In Section~\ref{sec:model} we introduce the
model of the studied quantum system. Section~\ref{sec:lindblad} shows different
approaches to solving the Lindblad equation. Next, in
Section~\ref{sec:gprogramming} we introduce genetic programming. Detailed
description of the optimization procedure and studied cases can be found in
Section~\ref{sec:optimizaiton}. In Section~\ref{sec:results} we show
results of the numerical simulations. Finally,  we summarize this work in
Section~\ref{sec:conclusions}.

\section{Model of the quantum system}\label{sec:model}

Our goal is to implement the unitary operations $U_\mathrm{NOT} = \sigma_x 
\otimes \1$ and $U_\mathrm{SWAP}= \1 \otimes \sum_{ij} \ketbra{i}{j} \otimes 
\ketbra{j}{i}$ on a quantum system 
modeled as a an isotropic Heisenberg spin-$1/2$ chain of 
a finite length $N$. We will study two and three qubit systems. The total 
Hamiltonian of the aforementioned quantum  control system is given by
\begin{equation}
	H(t) = H_0 + H_c(t),
\end{equation}
where 
\begin{equation}
	H_0 = J \sum_{i = 1}^{N - 1} 
	S_x^iS_x^{i+1}+S_y^iS_y^{i+1}+S_z^iS_z^{i+1},
\end{equation}
is a drift part given by the Heisenberg Hamiltonian. The control is performed
only on the $n^\textrm{th}$ spin and is Zeeman-like, i.e.
\begin{equation}
	H_c(t) = h_x(t)S_x^n + h_y(t)S_y^n.
\end{equation}
In the above $S_k^i$ denotes $k^{\text{th}}$ Pauli matrix acting on the spin 
$i$.
Time dependent control parameters $h_x(t)$ and $h_y(t)$ are chosen to be
piecewise constant. For notational convenience, we set $\hbar = 1$ and after 
this rescaling frequencies and control-field amplitudes can be expressed in 
units of the coupling strength $J$, and on the other hand all times can be 
expressed in units of $1/J$~\cite{heule2010local}.

We model the noisy quantum system using the Markovian approximation with the 
master equation in the Kossakowski-Lindlbad form
\begin{equation}
\frac{\dd\rho}{\dd t}=-\ii[H(t),\rho] + 
\sum_j \gamma_j(L_j\rho L_j^\dagger - 
\frac{1}{2}\{L_j^\dagger L_j,\rho\}),
\label{eq:master-equation}
\end{equation} 
where $L_j$ are the \emph{Lindblad} 
operators, representing the environment influence on the system
\cite{nielsen2000quantum} and $\rho$ is the state of the system.

The main goal of this paper is to compare various methods for optimizing 
control pulses $h_x(t), h_y(t)$ for the model introduced above.
In the next section we present three methods for this purpose.
The comparison of the control pulses obtained by different methods is done by 
applying these pulses into the above model and analysis of the obtained 
results.
\section{Various approaches to fidelity maximization}\label{sec:lindblad}
In this Section we describe three methods we used to Obtain optimal control 
pulses. The first method is an approximate method for obtaining a mapping 
which is close to a unitary one. The second uses an approximate derivative of 
the mapping with respect to control pulses.
Both of these methods allows to compute approximate gradient of the 
fitness function. In our numerical research we perform optimization with use 
of the L-BFGS-B optimization algorithm \cite{ zhu1997}. This 
algorithm requires efficient gradient computation. Its main advantages lies in 
harnessing the approximation  of Hessian of fitness function (we refer to the 
section 3 of \cite{byrd1994} for details).
Finally, we use genetic programming to optimize control pulses without the 
need for computing gradient of the fidelity function.

\subsection{Approximate method}
Assuming piecewise constant control pulses the Hamiltonian in 
Eq.~\eqref{eq:master-equation} becomes independent of time during the 
duration of the pulse. This allows us to simplify the master equation.

For notational convenience, let us write the decoherence part of the 
Eq.~\eqref{eq:master-equation} in the following form~\cite{havel2003robust}
\begin{equation}
-\GG = \sum_j \gamma_j \left(L_j \otimes \overline{L_j} - \frac{1}{2} \left[ 
\left( L_j^\dagger L_j \right) \otimes \1 + \1 \otimes 
\left(\overline{L_j^\dagger L_j}\right) \right] \right). \label{eq:decoherent}
\end{equation}

The second term in Eq.~\ref{eq:master-equation} can be written as
\begin{equation}
\ii\HH = \ii \left( \1 \otimes \overline{H} - H \otimes \1 
\right).\label{eq:coherent}
\end{equation}

This observations allow us to write a mapping representing the evolution of a 
system in an initial state $\rho$ under Eq.~\eqref{eq:master-equation} for 
time $t$ as
\begin{equation}
A = \exp(-t\FF),
\end{equation}
where $-\FF = -\GG - \ii\HH$. The final state of the evolution is
\begin{equation}
\mathrm{res}(\rho_f) = A \mathrm{res}(\rho),
\end{equation}
where $\mathrm{res}(\cdot)$ is a linear mapping defined as
\begin{equation}
\mathrm{res}(\ket{\phi}\bra{\psi}) = \ket{\phi}\ket{\psi}.
\end{equation}

We can approximate the superoperator $A$ as 
\begin{equation}
\begin{split}
A \approx & \\
 & \exp\left[ -\frac12 t \sum_j \left( L_j^\dagger L_j \right) \otimes 
\1 + \1 \otimes \left(\overline{L_j^\dagger L_j}\right) \right]
\times\\
\times& \exp\left( t \sum_j L_j \otimes \overline{L_j} \right) 
\times \exp \left( -t\ii\HH \right) + O(t^2)\\
=&\A(t)\BB(t)\CC(t) + O(t^2).
\end{split}
\end{equation}
Note that, only the $\CC(t)$ term depends on the control pulses. Assuming 
piecewise constant control pulses, we can write the resulting superoperator as
\begin{equation}
A = \prod_{i=1}^{n} \A(\Delta t_i)\BB(\Delta t_i)\CC(\Delta 
t_i),\label{eq:approx-evolution}
\end{equation}
where $n$ is the total number of control pulses and $\Delta t_i$ is the length 
of the time interval in which control pulse $h_i$ is applied to the system. 
The derivative of the superoperator with respect to a control pulse $h_l$ is
\begin{equation}
\begin{split}
	\frac{\partial A}{\partial h_l} = & \left( \prod_{i=1}^{l-1} \A(\Delta 
	t_i)\BB(\Delta t_i)\CC(\Delta t_i) \right) \times \\
	\times & \left(\A(\Delta t_l)\BB(\Delta t_l) \frac{\partial 
	\CC(\Delta t_l)}{\partial h_l} \right) \times \\
	\times & \left( \prod_{i=l+1}^{n} \A(\Delta t_i)\BB(\Delta t_i)\CC(\Delta 
	t_i) \right),
\end{split}\label{eq:superop-deriv}
\end{equation}

We use the fidelity as the figure of merit
\begin{equation}
f = \frac{1}{2^{2N}}\Re(\Tr A_\mathrm{T}^\dagger A),
\end{equation}
where $N$ is the number of qubits in the system and $A_\mathrm{T}$ is the
target superoperator. The derivative of the fidelity is given by
\begin{equation}
\frac{\partial f}{\partial h_l} = \frac{1}{2^{2N}} \Re\left( \Tr \left( 
A_\mathrm{T} \frac{\partial 
A}{\partial h_l} \right)\right).
\end{equation}
\subsection{Approximate gradient method}
In this section we follow the results by Machnes \emph{et. 
al.}~\cite{machnes2011comparing}.
In order to introduce the approximate gradient method, we introduce the 
following notation
\begin{equation}
\hat{H}(\cdot) = [H(t), \cdot].
\end{equation}
This allows us to write Eq.~\eqref{eq:master-equation} in the form
\begin{equation}
\frac{\partial \rho}{\partial t} = -(\ii\hat{H} + \LL)\rho(t).
\end{equation}
The evolution of a quantum map under these equation is given 
by
\begin{equation}
\frac{\partial X(t)}{\partial t} = -(\ii \hat{H} + 
\LL)X(t).\label{eq:superop-motion}
\end{equation}
In order to perform numerical simulations, Eq.~\eqref{eq:superop-motion} needs
to be discretized. Given a total evolution time $T$, we divide it into $M$
small intervals, each of length $\Delta t = T/M$. Hence, the quantum map in the
$k^\mathrm{th}$ time interval is given by
\begin{equation}
X_k = \exp\left[ -\Delta t  ( \ii \hat{H}(t_k) + \LL(t_k) ) \right].
\end{equation}

We utilize the trace fidelity as the figure of merit for this optimization 
problem
\begin{equation}
f = \frac{1}{2^{2N}}\Re\tr \left[ X^\dagger_\mathrm{target}X(T) \right] = 
\frac{1}{2^{2N}} 
\Re \tr \left[ \Lambda^\dagger(t_k)X(t_k) \right],
\end{equation}
where $X(t_k) = X_k X_{k-1} \cdots X_1 X_0$ and $\Lambda^\dagger(t_k) = 
X^\dagger X_M X_{M-1}\cdots X_{k+2} X_{k+1}$. This allows us to write the 
derivative of the fidelity with respect to the control pulses as
\begin{equation}
\frac{\partial f}{\partial h_j(t_k)} = \frac{1}{2^{2N}} \Re \tr \left[ 
\Lambda^\dagger(t_k) \left( \frac{\partial X_k}{\partial h_j(t_k)} \right) 
X(t_{k-1}) \right].
\end{equation}
Since $\LL$ and $\ii \hat{H}$ need not commute, we can not calculate the 
derivative $\frac{\partial X_k}{\partial h_j(t_k)}$ using exact methods. The 
best approach is to use the following approximation for the gradient
\begin{equation}
\frac{\partial X_k}{\partial h_j(t_k)} \approx -\Delta t \left( \ii \hat{H} + 
\frac{\partial \LL(h_j(t_k))}{\partial h_j(t_k)} \right) X_k.
\end{equation}
This approximation is valid provided that $$\Delta t \ll \frac{1}{||\ii 
\hat{H} 
+ \LL||_2}.$$

\subsection{Genetic programming}\label{sec:gprogramming}
Genetic programming (GP) is a numerical method based on the evolutionary
mechanisms  \cite{spector2004, gepp2009}. There are two main reasons for using 
GP for finding optimal
control pulses. First of all it enables to perform numerical search in
complicated, mathematically untraceable space.
On the other hand, one should note that
the values of control pulses in different time intervals can be set
independently. Thus the idea of genetic code fits well as a model for a
control setting.
Thanks to such an representation, genetic programming enables to exchange 
values of
control pulses in some fixed intervals between control settings that results
with maximally accurate approximation of the desired evolution.

\subsubsection{General GP algorithm}
Genetic programming belongs to the family of search heuristics inspired by the 
mechanism of natural evolution. 
Each element of a search space being candidate for a 
solution
is identified with a representative of a population.
Every member of a population has its unique genetic code, which
is its representation in optimization algorithm.
In most of the cases genetic code is a sequence of values from a fixed 
set $\Sigma$ of possible values of all the features that characterize a
potential solution 
$x\in\Sigma^n$ in the search space.
Searching for the optimal solution is done by the systematic modification and 
evaluation of genetic codes of population members 
due 
to the rules of
the evolution such as
mutation, selection, crossover and inheritance.

Mutators are functions that change single elements of a genetic code  
randomly. A basic example of a mutator is a function that randomly 
changes
values of a representative $x$ at all positions with some non-zero probability

\begin{equation}\label{eq::mutator}
M(x)_i = \left\{
\begin{array}{cc}
x_i,                   & \mathrm{probability}\hspace{0.1cm}p \\
\mathrm{rand}(\Sigma), & \mathrm{probability}\hspace{0.1cm}1-p
\end{array}
\right.\hspace{-0.2cm}.
\end{equation}

Crossovers implement the mechanism of inheritance.
This function divides given parental genetic
code and create a new genetic code.
Commonly two new codes are created at the same time from two parental codes.
An example of such crossover is so called two point cut, where both parental 
codes ($x_i, y_i$) are cut into three
regions and the middle segments are interchanged

\begin{equation}\label{eq::crossover}
x'_i = \left\{ \begin{array}{cc} x_i & i\le c_1 \lor c_2 \le i \\ y_i & 
c_1<i<c_2 \end{array}\right.\hspace{-0.15cm},
\hspace{0.5cm}
y'_i = \left\{ \begin{array}{cc} x_i & c_1<i<c_2 \\ y_i & i\le c_1 \lor c_2 
\le i \end{array} \right.\hspace{-0.15cm},
\end{equation}
where $c_1< c_2$ are randomly chosen indices.
In every iteration of the algorithm all members of the population are 
evaluated using fitness function $f:\Sigma^n \to \Re$
which enables elements ordering.
Then, using a selector function, the set of the best members is obtained and 
used 
to create a new generation of the population using mutation and crossover 
functions.
There is a number of strategies for defining selector function --
from completely random choices to the deterministic choice of best 
representatives.

Strategy based on evolution mechanism makes genetic programming especially 
usable
when parts of genetic code represent
features of elements of a search space that
can be interchanged between elements independently.
In such case GA is expected to find the features that occur in well fitted 
representatives
and mix them in order to find the best possible combination.
Pseudo code representing this approach is presented in Listing 1.

\begin{center}

\begin{minipage}{9cm}
\begin{verbatim}
population = RandomPopulation()
for( generationsNumber ){
    newPopulation = []
    for(i = 0; i<population.size()/2; i++){
        mom = Selector(population)
        dad = Selector(population)
        (sister, brother) = CrossOver(mom, dad)
        Mutator(sister)
        Mutator(brother)
        newPopulation.append(sister)
        newPopulation.append(brother)
    }
    population = newPopulation
}
\end{verbatim}
\label{code::gp}
\end{minipage}\\
{Listing 1: }%
Pseudo code representing the algorithm of genetic programming. Functions
{\texttt{ Selector, Mutator} and\texttt{ CrossOver}} work as defined in Section
\ref{sec:gprogramming}.
\end{center}

While the customization of population representation and fitness function 
unavoidably
relies on the optimization problem,
other parameters of genetic programming such as crossover and mutation methods
are universal.

\subsubsection{Customization of the GP}
Using GP schema requires obtaining proper representation of a problem.
First of all one need to model the space of possible solutions as a set of 
genomes, usually by representing each of unique and independent features of a 
solution as one gene. Secondly, it is necessary to define a fitness function 
that allows to estimate genomes in a way that is consistent with the 
optimization problem. Using this function one need to determine selection 
method. The last step is to define methods for modifying modeled genomes. It 
may be methods based on mutation, crossing-over or both.

In this work we investigate methods for optimizing the sequence of control 
pulses in order to perform given unitary evolution. As we assume that control 
pulses in each time step are independent, it is a natural to define each of 
subsequent control pulses as gene. In this case genome is modeled as sequence 
od real numbers. It is a very convenient method, because there are many 
already 
developed mutation and crossing-over methods for such genome model that have 
been successfully applied to the problems of searching for the optimal 
evolution \cite{goldberg1989, sadowski2013genetic}. Methods of selection do 
not depend on genome model and can be based on a variety of already existing 
ones as well.

Since GP schema does not require the ability to compute gradient of considered 
fitness function one can use any method for genome estimation.
In this work we use control pulses represented by given genome to perform 
simulation of the system evolution and obtain resulting state that is compared 
with the one resulting from target evolution. For comparison purposes we apply
functions described in section \ref{sec:optimizaiton}.

\subsubsection{Optimization}\label{sec:optimizaiton}
In order to optimize a controlled evolution of a system governed by the 
Lindblad equation, we perform optimization of the average of distances 
between 
target state operator and the resulting states for each basis matrix of the 
space of input states.
Our fitness function is defined as
\begin{equation}
f([c_0, ..., c_M]) = \frac{1}{2^{2N}} \sum_{i=1}^{2^{2N}} \Tr  
\rho^i_\mathrm{T} \Tr_A\left( 
\Phi( \rho^i_0, [c_0, ..., c_M] ) \right),
\end{equation}
where $\Tr_A$ denotes tracing out the ancila, $N$ is the total number of 
qubits in the system, $\rho^i_0$ denotes the $i^\mathrm{th}$ basis matrix, 
$\rho_\mathrm{T}^i = U_\mathrm{T}  \Tr_A(\rho_0^i)  U_\mathrm{T}^\dagger$ is 
the target density matrix and $\Phi$ is the quantum channel corresponding to 
the time evolution of $\rho_0^i$ under Eq.~\eqref{eq:master-equation} for 
control pulses $[c_0, ..., c_M]$.

We study two noise models, the amplitude damping and phase damping noise. The 
former is given by the Lindblad operator $L_1 = \sigma_- = \ketbra{0}{1}$, 
while the latter is $L_2 = \sigma_z$.

\section{Results and discussion}\label{sec:results}
In this section we present the results obtained for all of the methods 
introduced in section \ref{sec:lindblad}.
The comparison of the control pulses obtained by different methods is done by 
applying these pulses into the model from section \ref{sec:model} and 
analysis of the results.

In all of the simulations, we set the number of control pulses to 32 for
two-qubit systems and 128 for the three-qubit systems. We limit the strength of
the control pulses to $h_\mathrm{max} = 100$. In each case we split all of the
qubits forming the system into two subsystems: the one that performs some fixed
evolution and the auxiliary one.

To find the best spin chain configuration, we study the following systems:
\begin{enumerate}
\item One-qubit system with one-qubit ancilla. The 
control is performed on the target qubit.
\item Two-qubit system with no ancilla.
\item One-qubit system with one-qubit ancilla. The
control is performed on the ancillary qubit.
\item Three-qubit system with no ancilla.
\item Two-qubit system with one-qubit ancilla. The 
control is performed on the ancillary qubit.
\item Three-qubit system with no ancilla.
\end{enumerate}

\begin{figure}[!h]
\subfloat[]{\begin{tikzpicture}[domain=0:4] 

\path
(2,1) node[circle,inner sep=1pt,draw](q2) {0}
(0,1) node[circle,inner sep=1pt,draw](q4) {1};
\draw (q4) -- (q2) ;
\draw[->] (0,1.5) -- (q4) ;

\draw[decorate,decoration={brace,amplitude=3pt,mirror}] 
    (1.6,0.5)   -- (2.4,0.5);
\draw[decorate,decoration={brace,amplitude=3pt,mirror}] 
    (-0.4,0.5)   -- (0.4,0.5);

\node at (2.0,0.){auxiliary};
\node at (2.0,-0.5){system};
\node at (0.0,-0.0){$NOT$};
\end{tikzpicture}\label{fig:sys_a}}\hspace{0.5cm}
\subfloat[]{\begin{tikzpicture}[domain=0:4] 
\path
(2,1) node[circle,inner sep=1pt,draw](q2) {0}
(0,1) node[circle,inner sep=1pt,draw](q4) {1};
\draw (q4) -- (q2) ;
\draw[->] (2,1.5) -- (q2) ;

\draw[decorate,decoration={brace,amplitude=3pt,mirror}] 
    (-0.4,0.5)   -- (2.4,0.5);
\node at (2.0,0.0){$\1$};
\node at (0.0,0.0){$NOT$};
\node at (1.0,0.0){$\otimes$};
\end{tikzpicture}\label{fig:sys_b}}\\
\subfloat[]{\begin{tikzpicture}[domain=0:4] 
\path
(2,1) node[circle,inner sep=1pt,draw](q2) {0}
(0,1) node[circle,inner sep=1pt,draw](q4) {1};
\draw (q4) -- (q2) ;
\draw[->] (2,1.5) -- (q2) ;

\draw[decorate,decoration={brace,amplitude=3pt,mirror}] 
    (1.6,0.5)   -- (2.4,0.5);
\draw[decorate,decoration={brace,amplitude=3pt,mirror}] 
    (-0.4,0.5)   -- (0.4,0.5);

\node at (2.0,.0){auxiliary};
\node at (2.0,-0.5){system};
\node at (0.0,0.){$NOT$};
\end{tikzpicture}\label{fig:sys_c}}\hspace{0.5cm}
\subfloat[]{\begin{tikzpicture}[domain=0:4] 
\path
(0,0.5) node[circle,inner sep=1pt,draw](q1) {2}
(2,0) node[circle,inner sep=1pt,draw](q2) {0}
(2,1) node[circle,inner sep=1pt,draw](q3) {1};
\draw (q2) -- (q1) -- (q3) -- (q2);
\draw[->] (2,1.5) -- (q3) ;

\draw[decorate,decoration={brace,amplitude=3pt,mirror}] 
    (-0.4,-0.5)   -- (2.4,-0.5);

\node at (2.0,-1.){$\1\otimes\1$};
\node at (0.0,-1.){$NOT$};
\node at (1.0,-1.){$\otimes$};
\end{tikzpicture}\label{fig:sys_d}}\\
\subfloat[]{\begin{tikzpicture}[domain=0:4] 
\path
(0,0.5) node[circle,inner sep=1pt,draw](q1) {0}
(2,0) node[circle,inner sep=1pt,draw](q2) {1}
(2,1) node[circle,inner sep=1pt,draw](q3) {2};
\draw (q2) -- (q1) -- (q3) -- (q2);
\draw[->] (0,1) -- (q1) ;

\draw[decorate,decoration={brace,amplitude=3pt,mirror}] 
    (1.6,-0.5)   -- (2.4,-0.5);
\draw[decorate,decoration={brace,amplitude=3pt,mirror}] 
    (-0.4,-0.5)   -- (0.4,-0.5);

\node at (0.0,-1.){auxiliary};
\node at (0.0,-1.5){system};
\node at (2.0,-1.){$SW\hspace{-0.1cm}AP$};
\end{tikzpicture}\label{fig:sys_e}}\hspace{0.5cm}
\subfloat[]{\begin{tikzpicture}[domain=0:4] 
\path
(0,0.5) node[circle,inner sep=1pt,draw](q1) {0}
(2,0) node[circle,inner sep=1pt,draw](q2) {1}
(2,1) node[circle,inner sep=1pt,draw](q3) {2};
\draw (q2) -- (q1) -- (q3) -- (q2);
\draw[->] (0,1) -- (q1) ;

\draw[decorate,decoration={brace,amplitude=3pt,mirror}] 
    (-0.4,-0.5)   -- (2.4,-0.5);

\node at (2.0,-1.){$SW\hspace{-0.1cm}AP$};
\node at (0.0,-1.){$\1$};
\node at (1.0,-1.){$\otimes$};
\end{tikzpicture}\label{fig:sys_f}}
\caption{Systems used for numerical simulation}\label{fig::systems}
\end{figure}
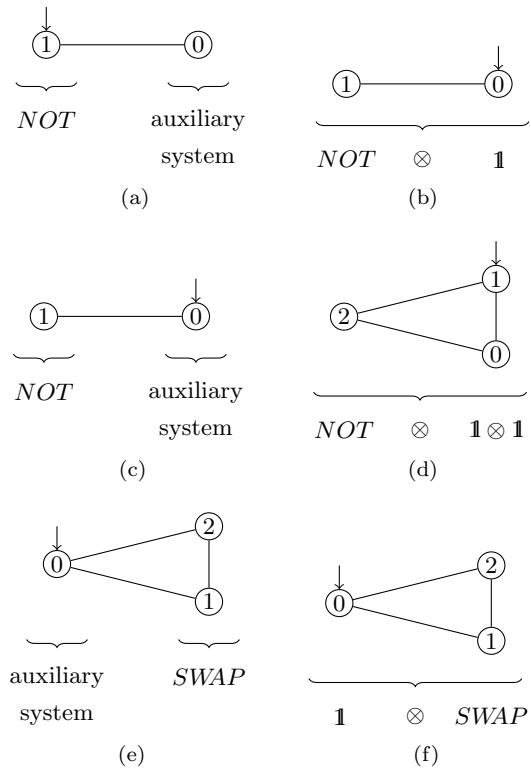

In our study we choose the phase damping and the amplitude damping channels as 
noise models. The former is given by the Lindblad operator $L=\sigma_z$, the 
latter is given by the operator $L=\sigma_- = \ketbra{0}{1}$. In order to 
objectively compare the algorithms, we study two setups: with equal number of 
steps of the algorithm and with equal computation time. The number of 
steps and length of these intervals are shown in Table~\ref{tab:deltas2} and 
Table~\ref{tab:deltas3}.

\begin{table}[!h]
\begin{tabular}{|l|p{1.5cm}|p{1.5cm}|p{1.5cm}|p{1.5cm}|p{1.5cm}|p{1.5cm}|}
\cline{2-7}
\multicolumn{1}{c}{} 
& \multicolumn{2}{|c|}{Genetic optimization}
& \multicolumn{2}{|c|}{Approximate evolution}
& \multicolumn{2}{|c|}{Approximate gradient} \\ \cline{2-7}
\multicolumn{1}{c|}{}
& $N$ & $\Delta t \; (10^{-3})$ & $N$ &$\Delta t \; (10^{-3})$ & $N$ & $\Delta 
t \; (10^{-3})$ \\ \hline
Equal number of steps & 32 & 65.625 & 32 & 65.625 & 32 & 65.625 \\ \hline
Equal computation time & 32 & 65.625 & 128 & 16.406 & 128 & 16.406 \\ \hline
\end{tabular}
\caption{Number of time steps and corresponding $\Delta t$ for different 
simulation setups in the two-qubit scenario}\label{tab:deltas2}
\end{table}

\begin{table}[!h]
\begin{tabular}{|l|p{1.5cm}|p{1.5cm}|p{1.5cm}|p{1.5cm}|p{1.5cm}|p{1.5cm}|}
\cline{2-7}
\multicolumn{1}{c}{} 
& \multicolumn{2}{|c|}{Genetic optimization}
& \multicolumn{2}{|c|}{Approximate evolution}
& \multicolumn{2}{|c|}{Approximate gradient} \\ \cline{2-7}
\multicolumn{1}{c|}{}
& $N$ & $\Delta t \; (10^{-3})$ & $N$ & $\Delta t \; (10^{-3})$ & $N$ & 
$\Delta t \; (10^{-3})$ \\ \hline
Equal number of steps & 128 & 16.406 & 128 & 16.406 & 128 & 16.406 \\ \hline
Equal computation time & 128 & 16.406 & 512 & 4.102 & 512 & 4.102\\ \hline
\end{tabular}
\caption{Number of time steps and corresponding $\Delta t$ for different 
simulation setups in the three-qubit scenario}\label{tab:deltas3}
\end{table}

\begin{figure}
\subfloat[One-qubit system with one-qubit ancilla. Target operation: NOT. The 
control is performed on the target qubit. Schematically shown in 
Figure~\ref{fig:sys_a}]
{\includegraphics{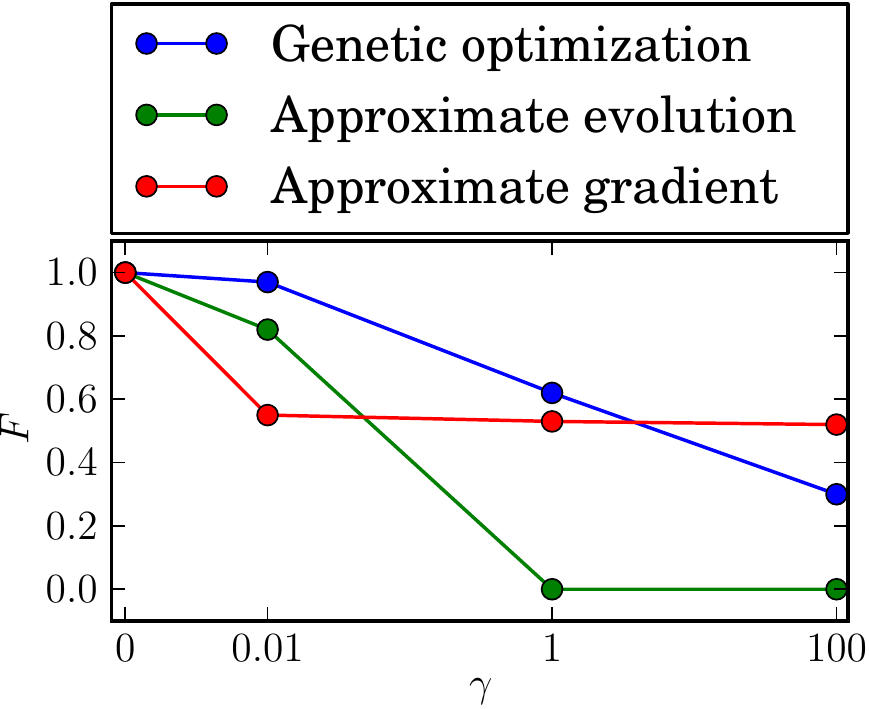}\label{fig:pd_a}}
\subfloat[Two-qubit system with no ancilla. Target operation: NOT. 
Schematically shown in Figure~\ref{fig:sys_b}]
{\includegraphics{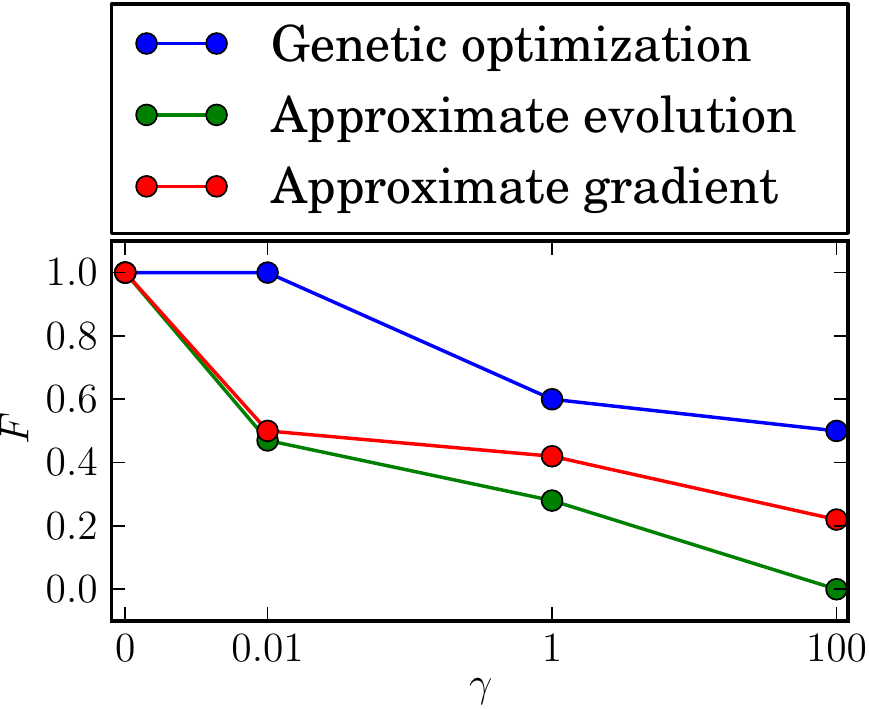}\label{fig:pd_b}}\\
\subfloat[One-qubit system with one-qubit ancilla. Target operation: NOT. The
control is performed on the ancillary qubit. Schematically shown in 
Figure~\ref{fig:sys_c}]
{\includegraphics{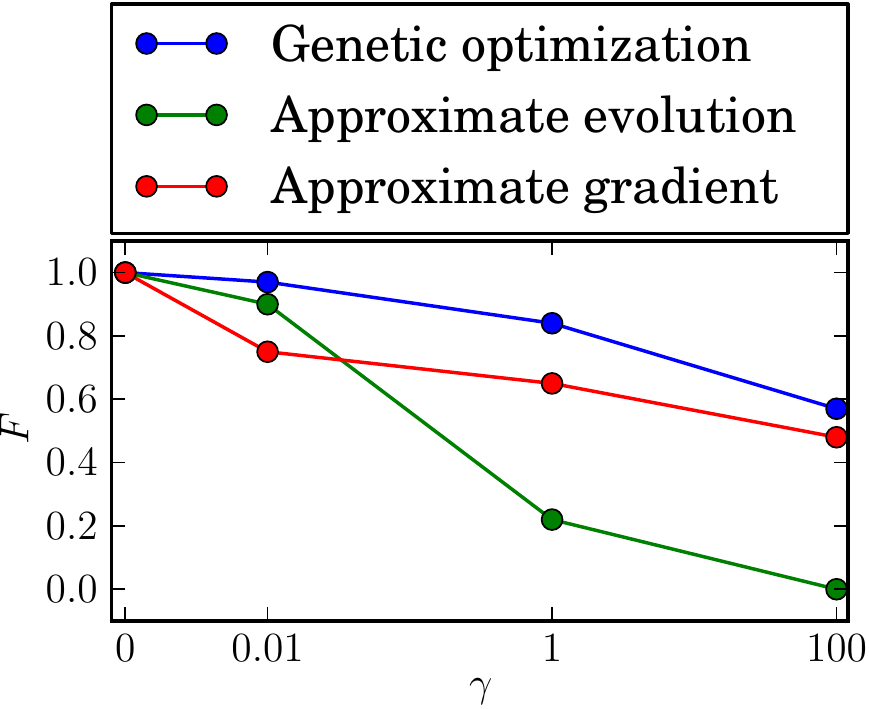}\label{fig:pd_c}}
\subfloat[Three-qubit system with no ancilla. Target operation: NOT. 
Schematically shown in Figure~\ref{fig:sys_d}]
{\includegraphics{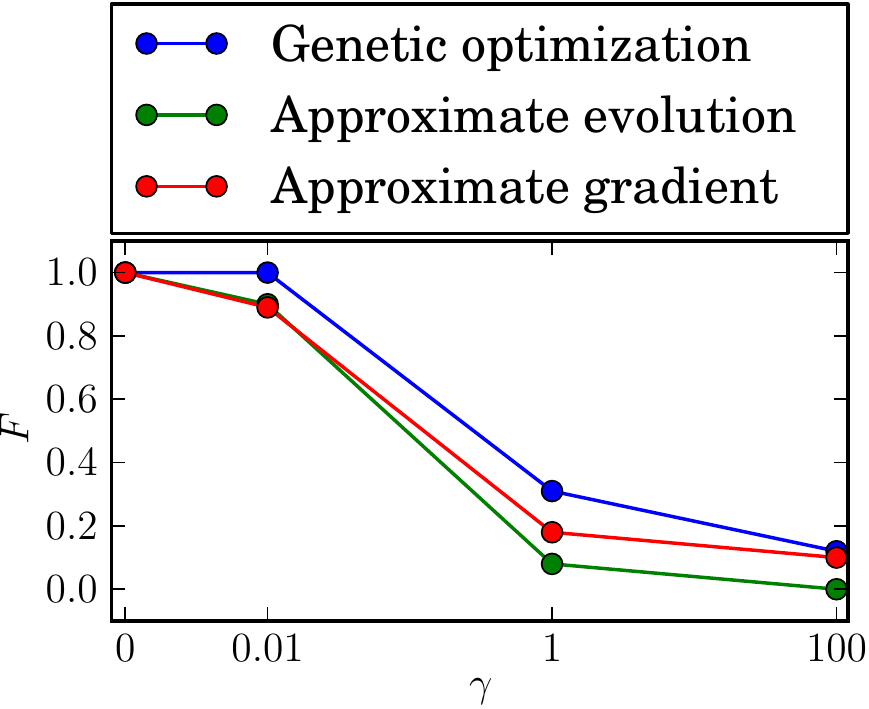}\label{fig:pd_d}}\\
\subfloat[Two-qubit system with one-qubit ancilla. Target operation: SWAP. The 
control is perfomed on the ancillary qubit. Schematically shown in 
Figure~\ref{fig:sys_e}]
{\includegraphics{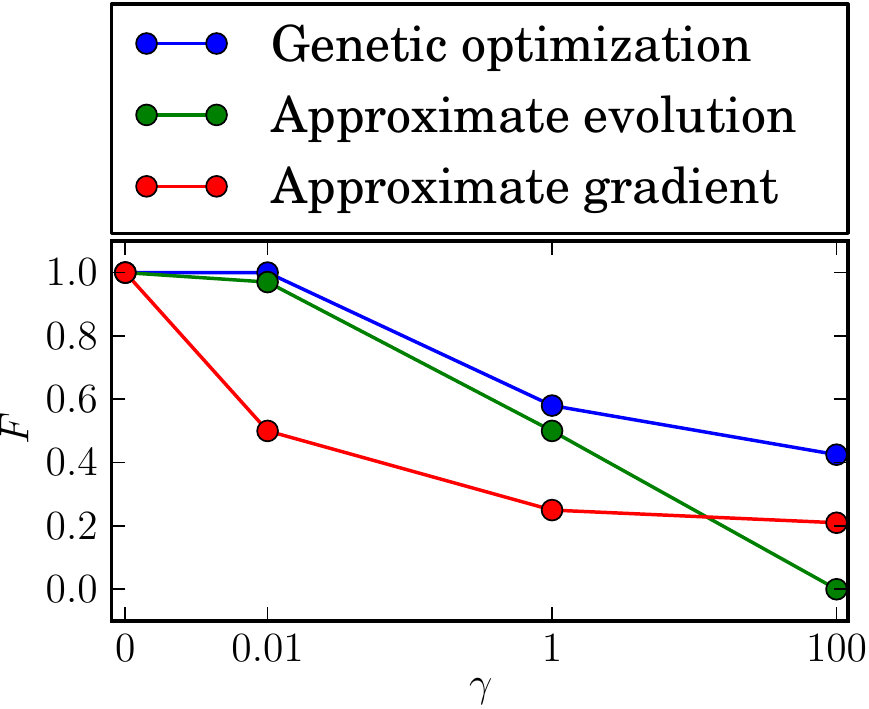}\label{fig:pd_e}}
\subfloat[Three-qubit system with no ancilla. Target operation: SWAP. 
Schematically shown in Figure~\ref{fig:sys_f}]
{\includegraphics{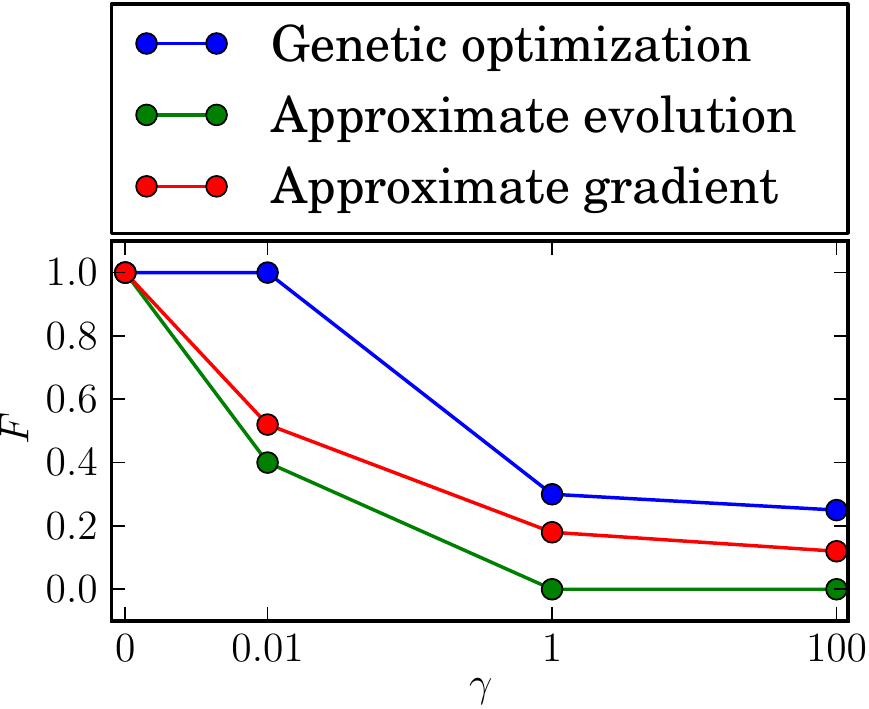}\label{fig:pd_f}}
\caption{Simulation results for the phase damping channel.}\label{fig:pd}
\end{figure}

Figure~\ref{fig:pd} shows the results for the phase damping channel. In this
case, the methods perform very similarly for the majority of studied cases. The
approximate evolution method tends to perform poorly for high noise values.
This is due to the fact that in this method, we have periods of coherent
evolution followed by decoherence, as Equation~\eqref{eq:approx-evolution}
states. This is in contrast with genetic optimization, where we make no
approximations and use the Lindblad equation. This results in simultaneous
decoherence and control. Thanks to this fact, the genetic optimization gives
the best results of all compared methods, even for high values of $\gamma$.
However, there is a trade-off. The computation time increases, at least,  by an
order of magnitude.

\begin{figure}
\subfloat[One-qubit system with one-qubit ancilla. Target operation: NOT. The 
control is performed on the target qubit. Schematically shown in 
Figure~\ref{fig:sys_a}]
{\includegraphics{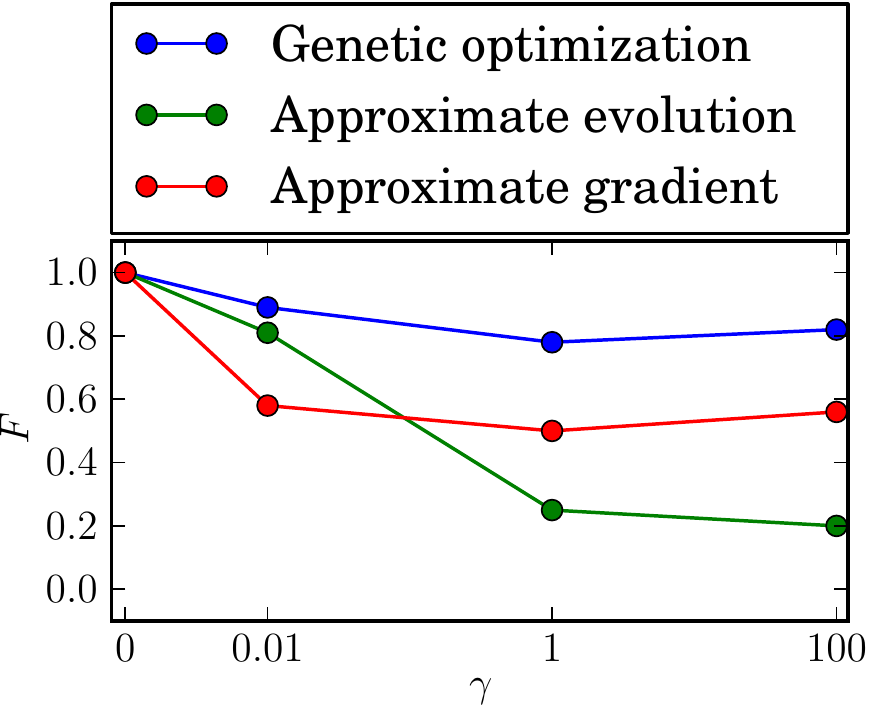}\label{fig:ad_a}}
\subfloat[Two-qubit system with no ancilla. Target operation: NOT. 
Schematically shown in Figure~\ref{fig:sys_b}]
{\includegraphics{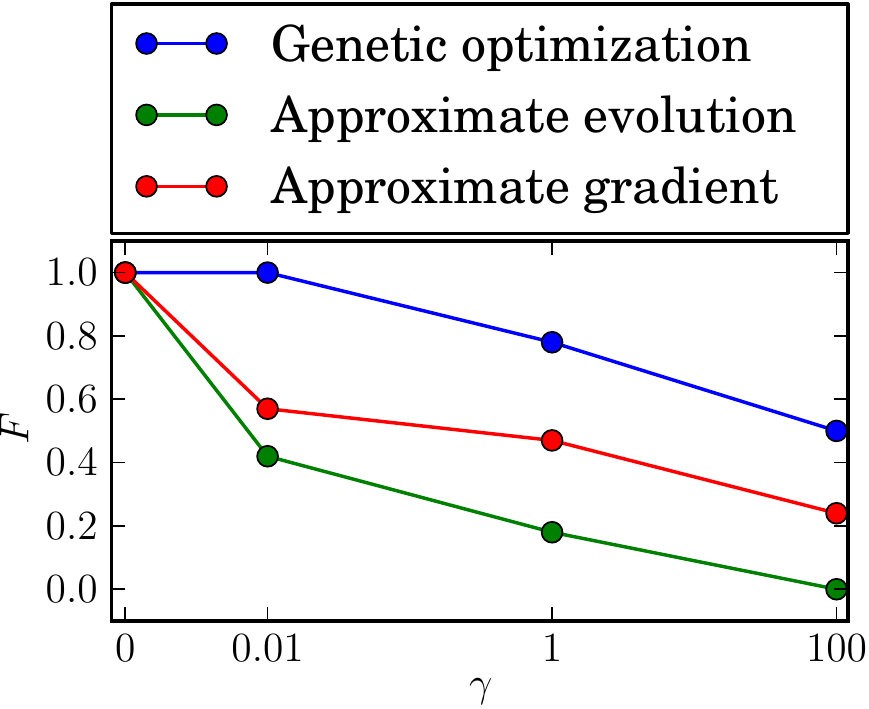}\label{fig:ad_b}}\\
\subfloat[One-qubit system with one-qubit ancilla. Target operation: NOT. The
control is performed on the ancillary qubit. Schematically shown in 
Figure~\ref{fig:sys_c}]
{\includegraphics{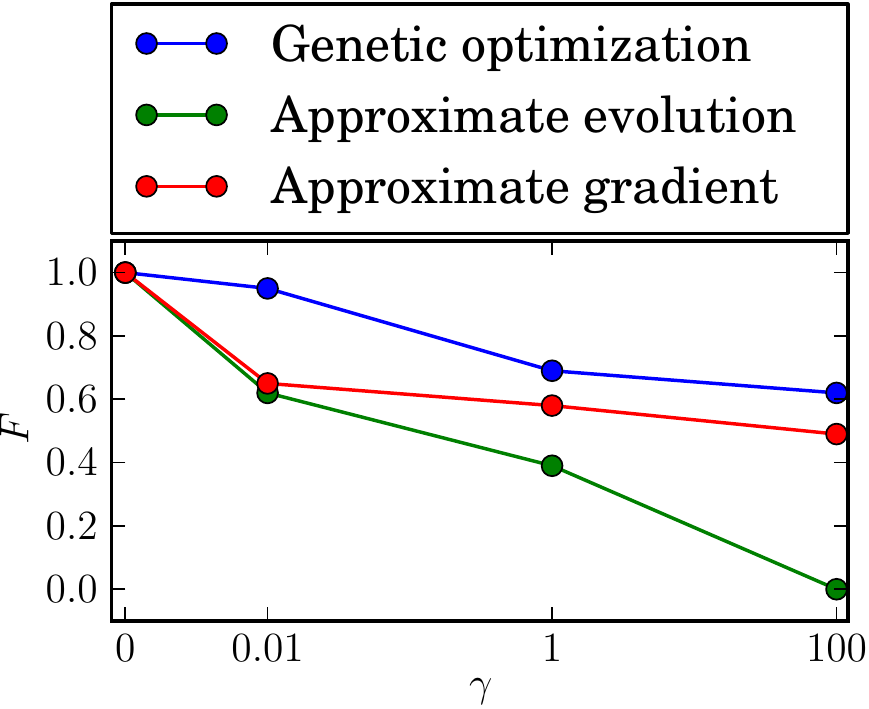}\label{fig:ad_c}}
\subfloat[Three-qubit system with no ancilla. Target operation: NOT. 
Schematically shown in Figure~\ref{fig:sys_d}]
{\includegraphics{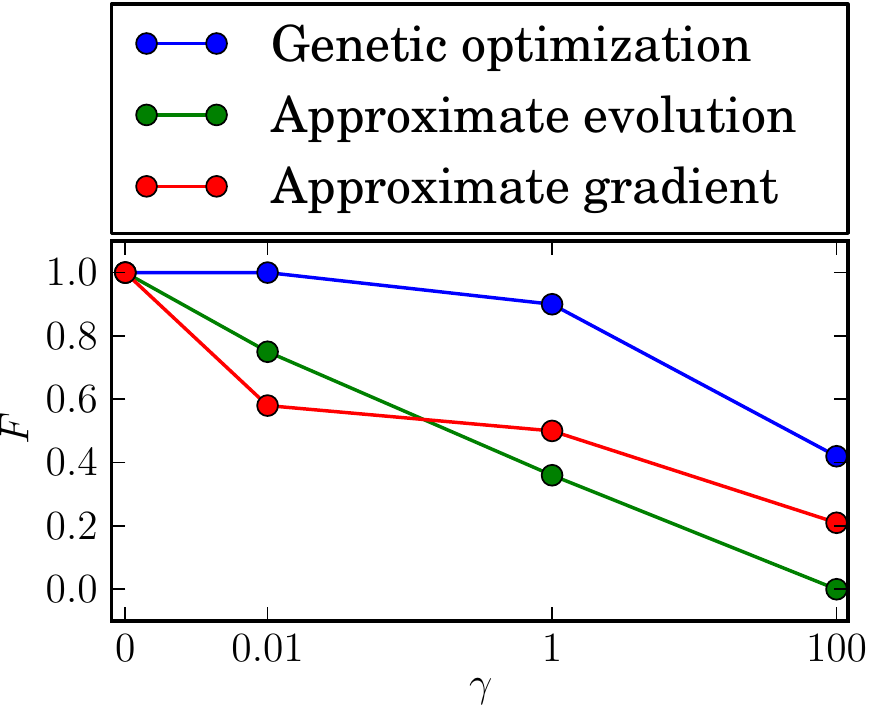}\label{fig:ad_d}}\\
\subfloat[Two-qubit system with one-qubit ancilla. Target operation: SWAP. The 
control is perfomed on the ancillary qubit. Schematically shown in 
Figure~\ref{fig:sys_e}]
{\includegraphics{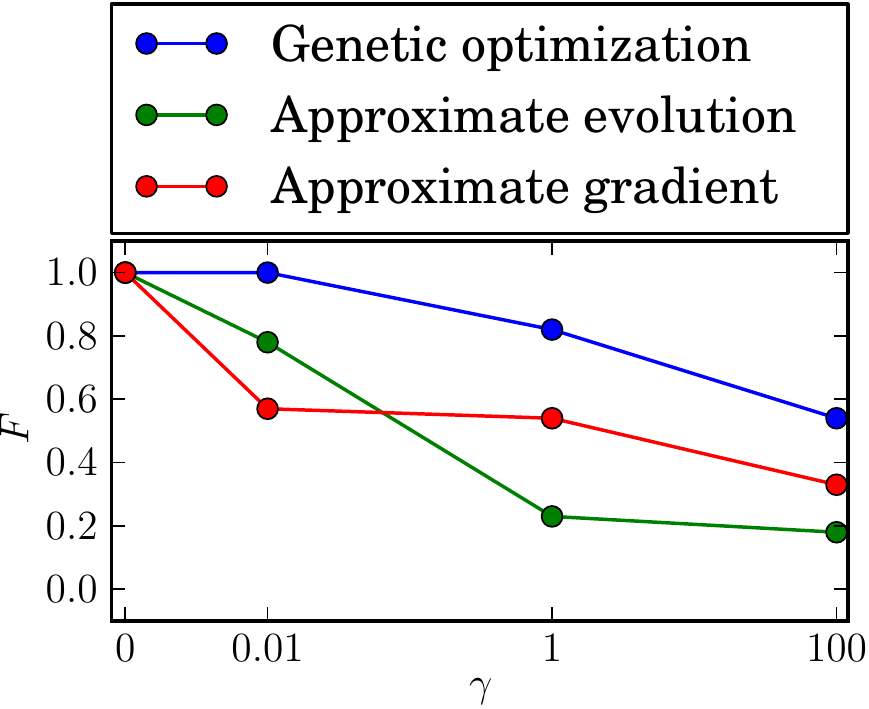}\label{fig:ad_e}}
\subfloat[Three-qubit system with no ancilla. Target operation: SWAP. 
Schematically shown in Figure~\ref{fig:sys_a}]
{\includegraphics{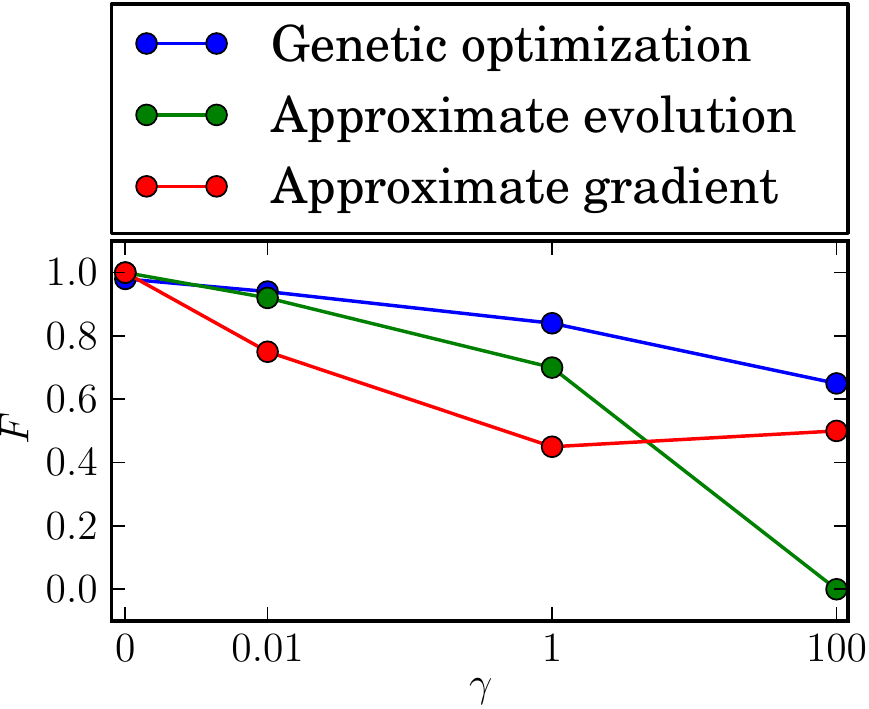}\label{fig:ad_f}}
\caption{Simulation results for the amplitude damping channel.}\label{fig:ad}
\end{figure}

Next, in Figure~\ref{fig:ad} we show the results for the amplitude damping 
channel. Similarly to the phase damping case, the approximate evolution method 
performs the worst. The approximate gradient method gives far better results, 
especially for high values of $\gamma$. Again, this may be explained by the 
fact that in the approximate evolution, we have periods of coherent evolution, 
followed by decoherence. On the other hand, the genetic optimization performs 
quite well, even for high noise values we were able to find sets of control 
parameters which gave fidelity higher than a half. Again, the trade-off was 
the computation time. Genetic optimization took about an order of magnitude 
longer compared to other methods.

Unfortunately, all the studied methods appear to fail for $\gamma > 0.01$ in 
the majority of investigated cases. For $\gamma=0.01$ only the genetic 
optimization gives a high value of fidelity.

\begin{figure}
\subfloat[One-qubit system with one-qubit ancilla. Target operation: NOT. The 
control is performed on the target qubit. Schematically shown in 
Figure~\ref{fig:sys_a}]
{\includegraphics{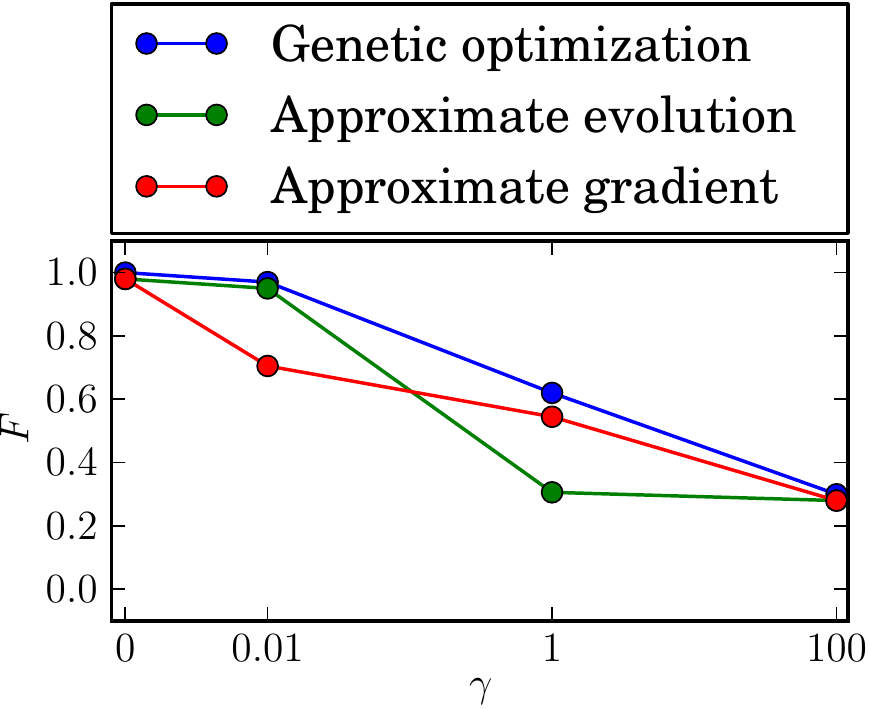}\label{fig:equal-time-phase_a}}
\subfloat[Two-qubit system with no ancilla. Target operation: NOT. 
Schematically shown in Figure~\ref{fig:sys_b}]
{\includegraphics{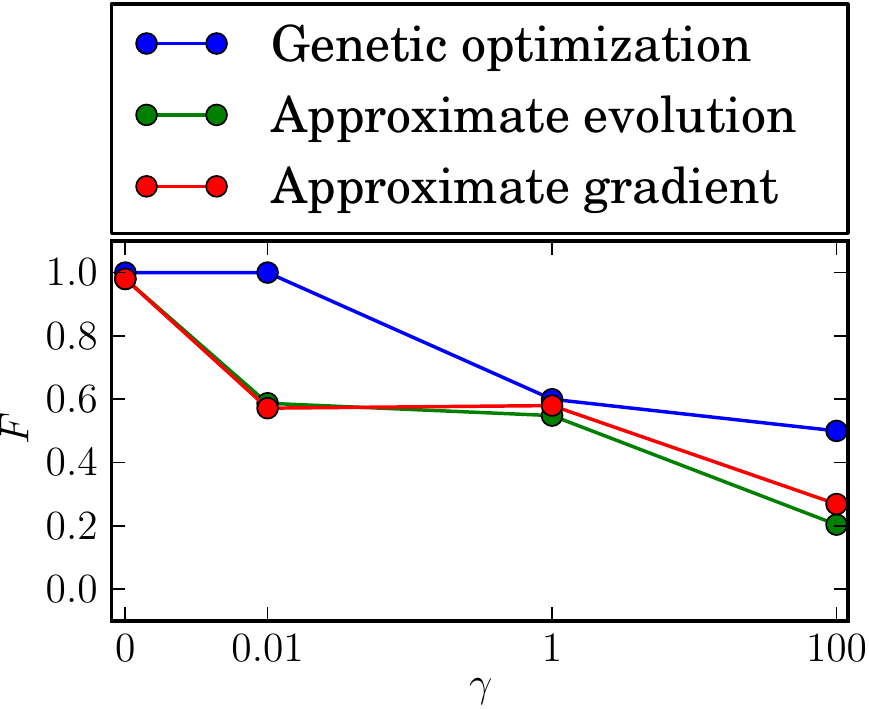}\label{fig:equal-time-phase_b}}\\
\subfloat[One-qubit system with one-qubit ancilla. Target operation: NOT. The
control is performed on the ancillary qubit. Schematically shown in 
Figure~\ref{fig:sys_c}]
{\includegraphics{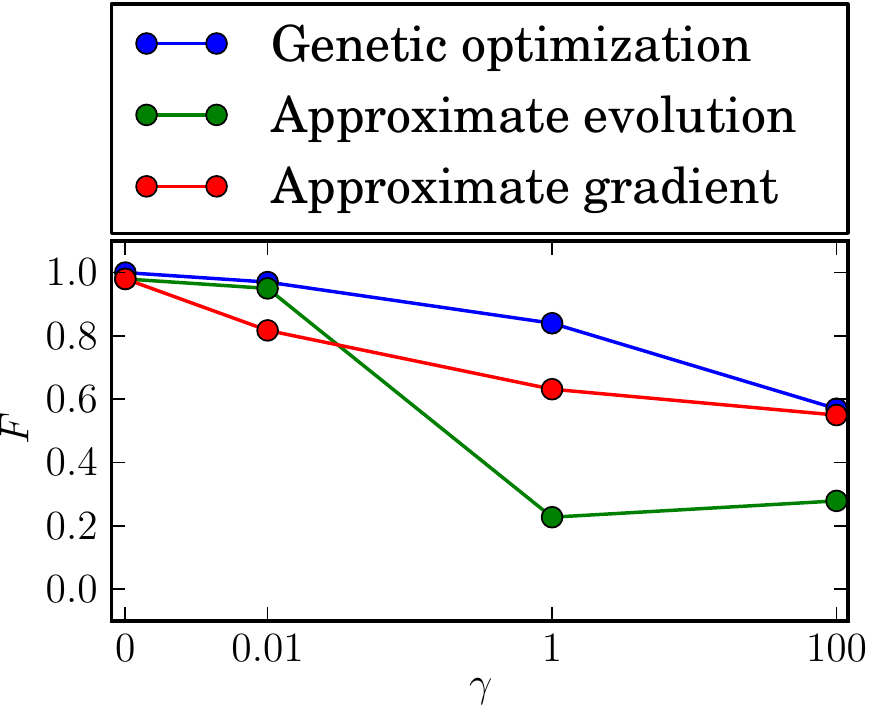}\label{fig:equal-time-phase_c}}
\subfloat[Three-qubit system with no ancilla. Target operation: NOT. 
Schematically shown in Figure~\ref{fig:sys_d}]
{\includegraphics{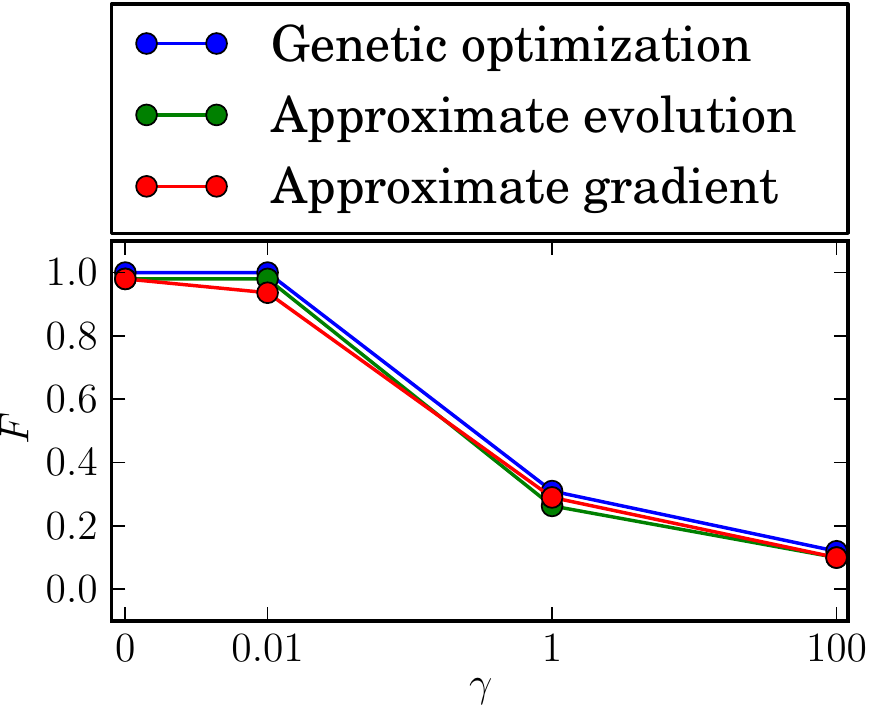}\label{fig:equal-time-phase_d}}\\
\subfloat[Two-qubit system with one-qubit ancilla. Target operation: SWAP. The 
control is performed on the ancillary qubit. Schematically shown in 
Figure~\ref{fig:sys_e}]
{\includegraphics{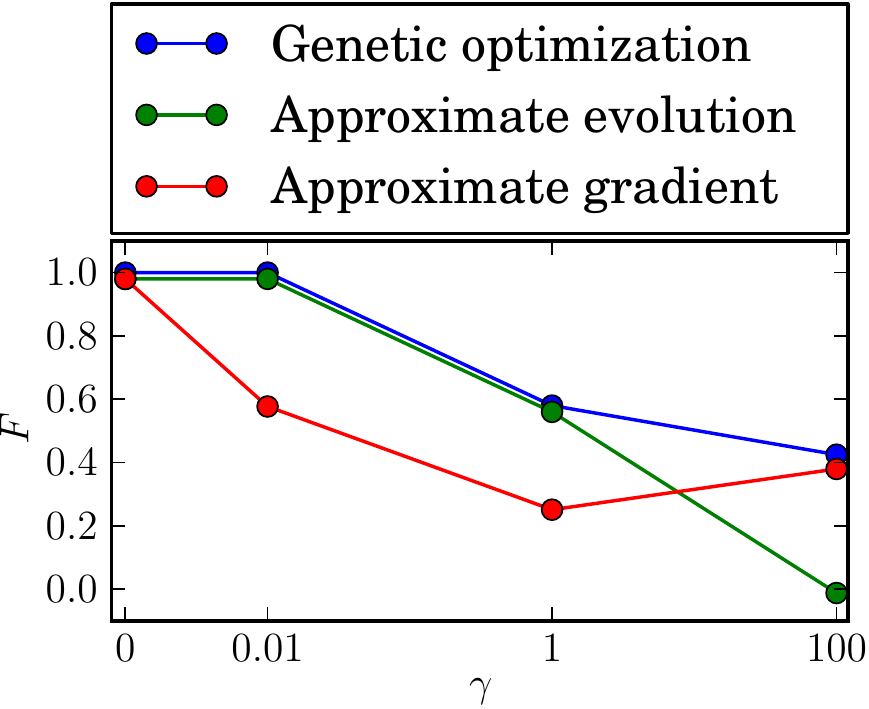}\label{fig:equal-time-phase_e}}
\subfloat[Three-qubit system with no ancilla. Target operation: SWAP. 
Schematically shown in Figure~\ref{fig:sys_f}]
{\includegraphics{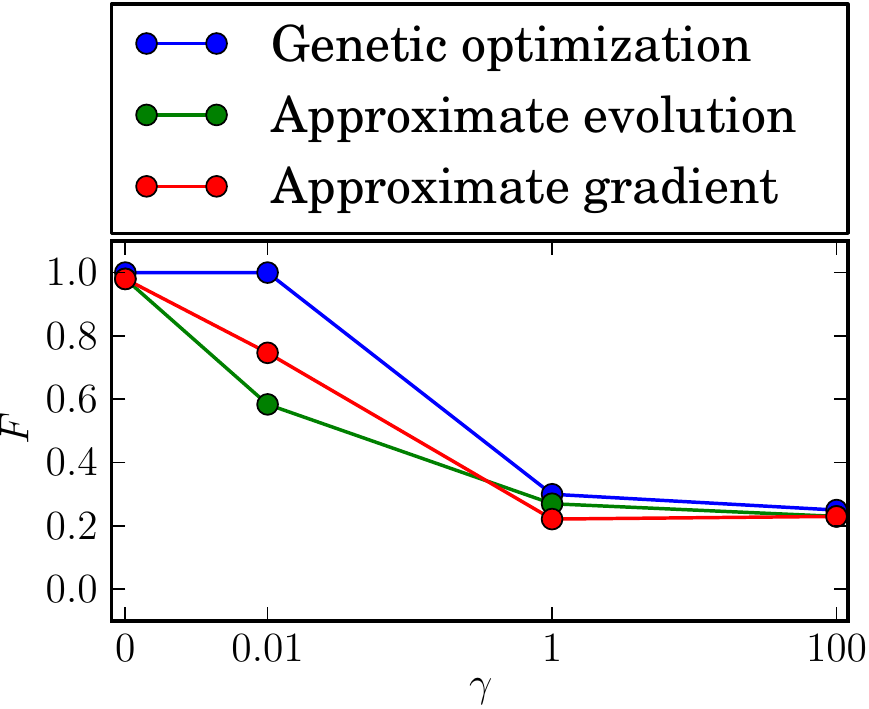}\label{fig:equal-time-phase_f}}
\caption{Simulation results for the phase damping channel with equal 
computation time.}\label{fig:equal-time-phase}
\end{figure}

\begin{figure}
\subfloat[One-qubit system with one-qubit ancilla. Target operation: NOT. The 
control is performed on the target qubit. Schematically shown in 
Figure~\ref{fig:sys_a}]
{\includegraphics{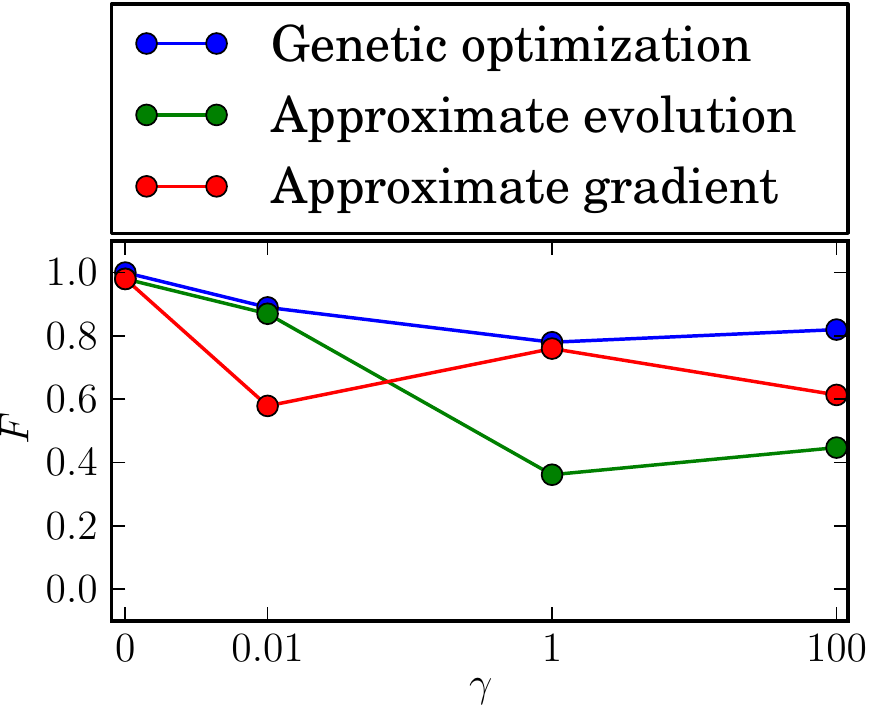}\label{fig:equal-time-amplitude_a}}
\subfloat[Two-qubit system with no ancilla. Target operation: NOT. 
Schematically shown in Figure~\ref{fig:sys_b}]
{\includegraphics{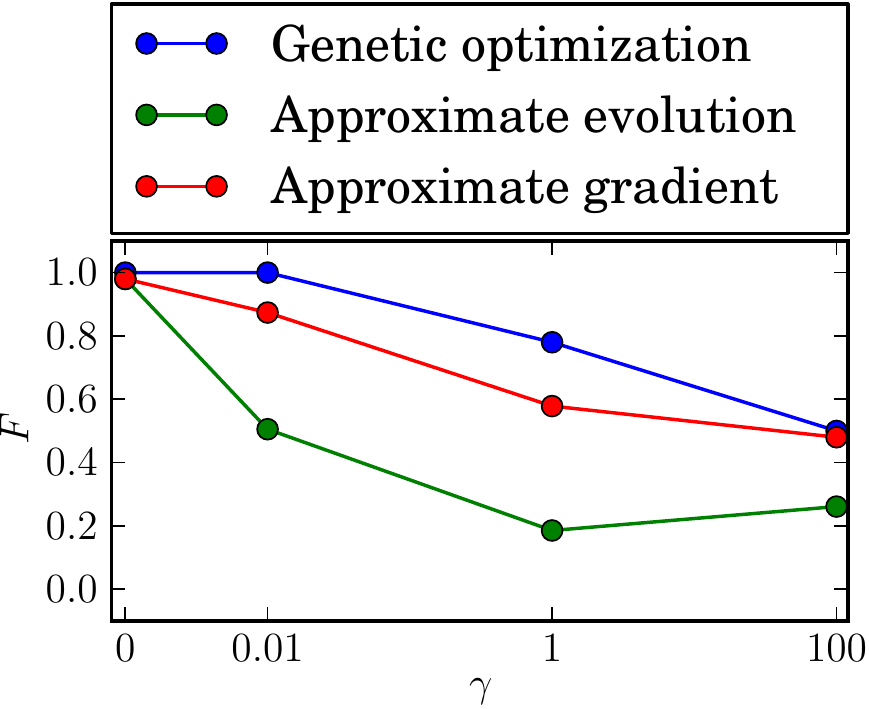}\label{fig:equal-time-amplitude_b}}\\
\subfloat[One-qubit system with one-qubit ancilla. Target operation: NOT. The
control is performed on the ancillary qubit. Schematically shown in 
Figure~\ref{fig:sys_c}]
{\includegraphics{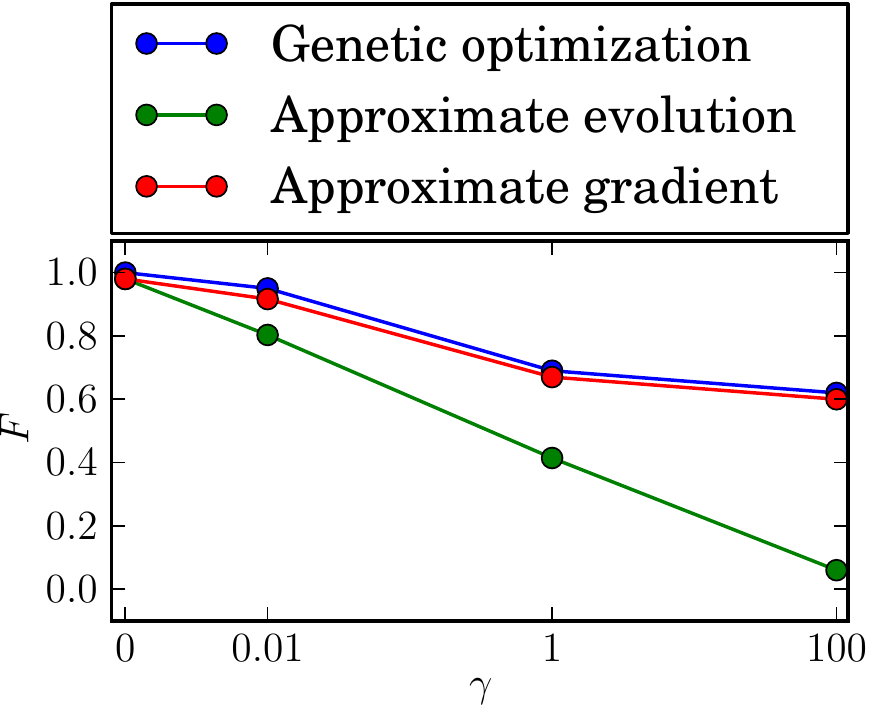}\label{fig:equal-time-amplitude_c}}
\subfloat[Three-qubit system with no ancilla. Target operation: NOT. 
Schematically shown in Figure~\ref{fig:sys_d}]
{\includegraphics{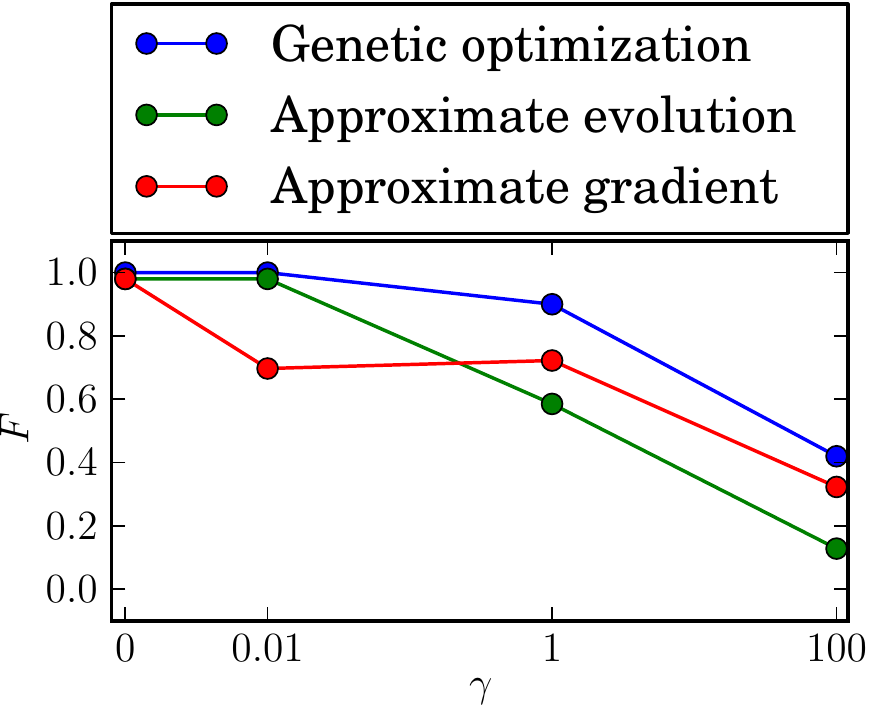}\label{fig:equal-time-amplitude_d}}\\
\subfloat[Two-qubit system with one-qubit ancilla. Target operation: SWAP. The 
control is performed on the ancillary qubit. Schematically shown in 
Figure~\ref{fig:sys_e}]
{\includegraphics{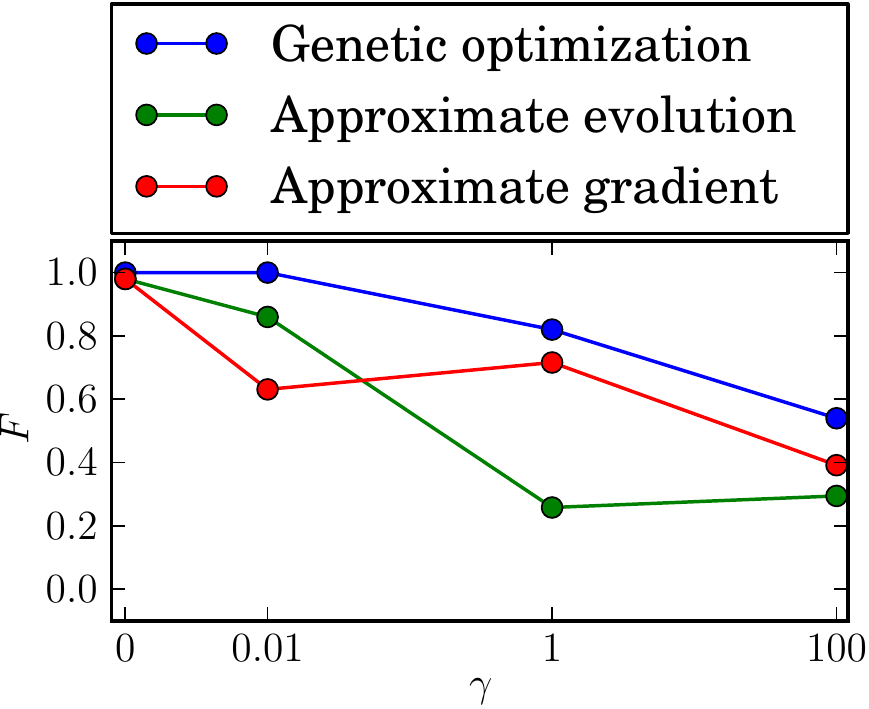}\label{fig:equal-time-amplitude_e}}
\subfloat[Three-qubit system with no ancilla. Target operation: SWAP. 
Schematically shown in Figure~\ref{fig:sys_a}]
{\includegraphics{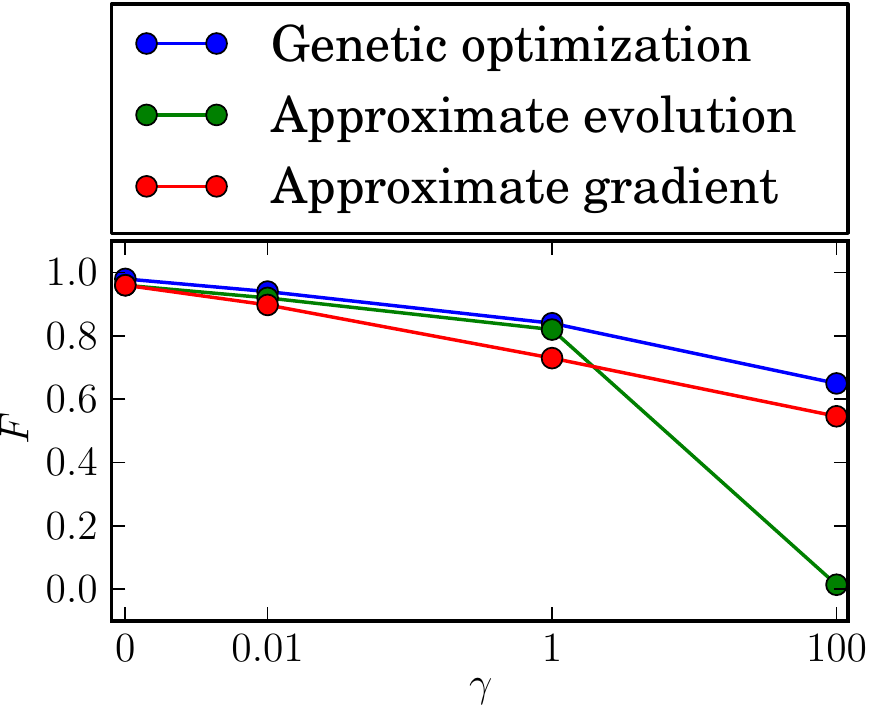}\label{fig:equal-time-amplitude_f}}
\caption{Simulation results for the amplitude damping 
channel with equal computation time.}\label{fig:equal-time-amplitude}
\end{figure}

The next step of our study focuses on comparing the algorithms when the
computation times are on the same order of magnitude. To achieve this, we added
more control pulses in the gradient based methods. In the case of both gradient
based methods we used 128 pulses for the two-qubit scenario and 512 for the
three-qubit scenario. The computation times are summarized in
Table~\ref{tab:computation}. Results are presented in
Figures~\ref{fig:equal-time-phase} and \ref{fig:equal-time-amplitude}. Also in
this setup the genetic optimization performs better compared to gradient based
approaches. In this case the gap between these methods is narrower than in the
case with equal number of control pulses.

\begin{table}[!h]
\begin{tabular}{|l|p{1.5cm}|p{1.5cm}|p{1.5cm}|p{1.5cm}|p{1.5cm}|p{1.5cm}|}
\cline{2-7}
\multicolumn{1}{c}{} 
& \multicolumn{2}{|c|}{Genetic optimization}
& \multicolumn{2}{|c|}{Approximate evolution}
& \multicolumn{2}{|c|}{Approximate gradient} \\ \hline
Number of qubits
& 2 & 3 & 2 & 3 & 2 & 3 \\ \hline
Equal number of steps & 810 & 30254 & 824 & 4147 & 781 & 3943 \\ \hline
Equal computation time & 8241 & 31029 & 8101 & 30846 & 8128 & 30447 \\ \hline
\end{tabular}
\caption{Average computation times in seconds for different simulation 
setups.}\label{tab:computation}
\end{table}

\section{Conclusions}\label{sec:conclusions}

We studied different methods of obtaining piecewise constant control pulses
that implement an unitary evolution on a system governed by
Kossakowski-Lindblad equation. The studied methods included genetic
optimization and the BFGS algorithm with the use of fidelity gradient 
based on an approximate evolution of the quantum system and an
approximate gradient method for the exact evolution case. Our results show
that, by adding an ancilla, it is possible to implement a unitary evolution on
a system under the Markovian approximation. Furthermore, the results heavily
depend not only on the size, but also on the location of the ancilla in the 
spin chain.

What one can notice about the possibility to perform unitary computation in 
noisy quantum systems is that in majority of the cases it is much better to 
treat non-target qubits as an ancilla. When comparing systems extended with 
auxiliary qubits (labeled as $a,c,e$) and systems with no ancilla ($b,d,f$) 
the difference in possible approximation is emphatic.

The genetic optimization method outperforms gradient based methods in two 
studied setups: with equal number of control pulses and with equal computation 
time.

\section*{Acknowledgements}
Work by {\L}ukasz Pawela was supported by the Polish National Science Centre 
under the grant number DEC-2012/05/N/ST7/01105. Przemys{\l}aw Sadowski was 
supported by the Polish National Science Centre under the grant number 
N N514 513340.

\bibliography{superoperator_control}
\bibliographystyle{apsrev}
\end{document}